%% file: main.tex
\documentclass[sn-mathphys-num,iicol]{sn-jnl}
\usepackage{float}
\usepackage{tikz}
\usepackage[T1]{fontenc}
\usepackage{amsmath} 
\usepackage[compat=1.1.0]{tikz-feynman}
\usetikzlibrary{decorations.pathmorphing,patterns,calc,math}

\input{tikz}
\begin{document}

\title{Search for charged long-lived particles at the Future Circular Collider (FCC-ee)}

\author[a,b]{Soumyaa Vashishtha}
\author[c]{Mirindra Randriamanantena}
\author[d]{Thomas Creuset}
\author[a]{Maximilian Goblirsch-Kolb}
\author[a]{Isabell Melzer-Pellmann}

\affil*[a]{\orgname{Deutsches Elektronen-Synchrotron DESY},
\orgaddress{\street{Notkestr. 85}, \postcode{22607}, \city{Hamburg},
\country{Germany}}}

\affil[b]{\orgname{Institut für Experimentalphysik, Universität Hamburg},
\orgaddress{\street{Luruper Chaussee 149}, \postcode{22761}, \city{Hamburg},
\country{Germany}}}

\affil[c]{\orgname{University of Antananarivo},
\orgaddress{\country{Madagascar}}}

\affil[d]{\orgname{École Nationale Supérieure de Techniques Avancées (ENSTA)},
\orgaddress{\street{828 Boulevard des Maréchaux}, \postcode{91762} \city{Palaiseau Cedex}, \country{France}}}

\email{soumyaa.vashishtha@desy.de}
\abstract{\input{abstract}}
\keywords{Long-lived particles, FCC-ee, kinked tracks, time-of-flight}

\maketitle

\section{Introduction}
The standard model (SM) is a successful quantum field theory that describes the fundamental particles and three of the four fundamental interactions~\cite{Particle_physics_review}. It has been extensively tested at TeV energy scales at the LHC and Tevatron. But it is an incomplete model as it does not account for dark matter and dark energy, which contribute up to 95\% of the total energy density of the universe~\cite{DM_evidence_1}. The SM also does not include a quantum theory of gravity. Various beyond the standard model (BSM) theories, like supersymmetry (SUSY), provide solutions to the current limitations of the SM by predicting new particles~\cite{MARTIN_1998}. Many of these predict massive long-lived particles (LLPs), which travel macroscopic distances before decaying. \newline
Neutral LLPs may escape the detector, leaving only missing energy or isolated energy deposits, whereas charged LLPs can produce tracks with atypical ionization ($dE/dx$) or delayed time-of-flight. The experimental signatures of LLPs depend strongly on their proper decay length, $c\tau$~\cite{LLP_alimena} where $c$ is the speed of light and $\tau$ is the particle's lifetime. Decays occurring inside the tracking volume can lead to kinked, disappearing, or displaced tracks.

Although proposed e+e- colliders are primarily conceived as precision facilities, their clean collision environment and the high-performance tracking systems of their proposed detectors also provide a powerful opportunity for direct searches for new particles. In particular, charged long-lived particles decaying inside the tracker can produce kinked tracks, a distinctive signature that is difficult to reconstruct in the high-occupancy environment of hadron colliders \cite{Asai2011, Klamka2026}. The detectors feature timing layers whose information can also be exploited to study detector-stable LLPs that travel slower than SM particles, probing regions of the LLP parameter space, particularly at longer lifetimes, that remain largely unexplored by current and previous experiments.


As a benchmark in this study, we use long-lived staus $\tilde{\tau}$, the supersymmetric partners of the tau lepton, as predicted in the Minimal Supersymmetric Standard Model (MSSM) with gauge-mediated supersymmetry breaking~(GMSB)~\cite{GMSB_ee, GMSB_basic, GMSBllp, japan_gmsb}. We study, for the first time at FCC-ee, the sensitivity of the IDEA detector concept to stau pair production, combining kinked-track, displaced-vertex, and detector-stable signatures to cover a broad range of stau lifetimes. The analysis is performed within the \textsc{FCCAnalyses} framework~\cite{theideastudygroup2025ideadetectorconceptfccee,FCCAnalyses}.

\section{Future Circular Collider} 
The Future Circular Collider (FCC) is proposed to be constructed at CERN, succeeding the current LHC program~\cite{FeasibilityReport}. It will run in two phases: the FCC-ee delivering lepton collisions and the FCC-hh phase targeting hadron-hadron collisions. The FCC-ee is designed to operate at several center-of-mass energies corresponding to key physics thresholds: the Z pole ($\sqrt{s}= 91$~GeV), the WW production threshold ($\sqrt{s} = 160$~GeV), the Higgsstrahlung (ZH) ($\sqrt{s} = 240$~GeV), and the top-quark pair production threshold ($\sqrt{s} = 365$~GeV). This study focuses on simulated electron\text{--}positron collisions at $\sqrt{s} = 240$~GeV and $\sqrt{s} = 365$~GeV, using the Innovative Detector for $e^{+}e^{-}$ Accelerators (IDEA) detector concept, which provides high-precision tracking, vertex reconstruction and timing information well suited for long-lived particle searches. 

\subsection{IDEA detector at FCC}
The design of the IDEA detector emphasizes precise tracking, timing, and calorimetry, making it particularly well-suited for searches for charged LLPs \cite{theideastudygroup2025ideadetectorconceptfccee}. The innermost component is a silicon vertex detector, which consists of an inner and outer layer made of monolithic active pixel sensors. The inner layer extends from 13.7~mm to 35.6~mm from the beamline, while the outer layer covers 130\text{--}315~mm. This detector provides high-resolution vertexing, essential for identifying displaced decays. Surrounding the vertex detector is a lightweight drift chamber, enclosed within a silicon wrapper covering more than 100~m$^2$. The combination of drift chamber and silicon wrapper provides precise tracking and timing information over a radial distance of approximately 2~m, enabling accurate reconstruction of charged particle trajectories before they enter the calorimeter system. The electromagnetic calorimeter (ECAL) is a dual-readout crystal calorimeter, offering excellent energy resolution and a timing resolution of roughly 30~ps for electromagnetic showers above 30~GeV. The ECAL and tracker are enclosed by a 3~T solenoidal magnet. Outside the solenoid, the hadronic calorimeter (HCAL) employs a dual-readout design to achieve high-precision measurements of hadronic showers. The muon system forms the outermost layer, providing robust muon identification and momentum reconstruction.

\subsection{Signal Samples}
The GMSB is used as a benchmark model in this study where the stau~($\tilde{\tau}$), the superpartner of the lepton is the LLP. The stau decays to a gravitino~($\tilde{G}$) and a tau lepton~($\tau$). Signal events are generated using \textsc{MadGraph5\_aMC@NLO} v3.2.0~\cite{madgraph} for unpolarized $e^+ e^-$ collisions at $\sqrt{s} = 240$~GeV and $\sqrt{s} = 365$~GeV using the GMSB model \cite{Benjamin_Fuks}. The decay $\tilde{\tau}\rightarrow \tau \tilde{G}$ is enforced with a branching ratio of $100\%$ with each signal sample containing $10^5$ events. Parton-level events are then passed to \textsc{Pythia8}~\cite{pythia8} for parton showering and hadronization. Initial- and final-state radiation effects are included to model realistic kinematics.

The backgrounds shown in Table~\ref{background_samples} are chosen based on their large production rates and their potential to mimic the signal topology.

\begin{table*}[htbp]
\centering
\caption{Background samples considered at $\sqrt{s}=240$~GeV (left) and $\sqrt{s}=365$~GeV (right).}
\label{background_samples}

\begin{minipage}[t]{0.48\linewidth}
\centering
\textbf{$\sqrt{s}=240$~GeV}

\vspace{0.5em}

\begin{tabular}{ll}
\hline
Process & Decay \\
\hline
$e^+ e^- \to W^+ W^-$ & Inclusive $WW$ \\
$e^+ e^- \to ZZ$ & Inclusive $ZZ$ \\
$e^+ e^- \to \tau^+ \tau^-$ & Inclusive $\tau^+\tau^-$ \\
$e^+ e^- \to ZH$ & $H \to b\bar{b}$ \\
$e^+ e^- \to \nu_e \bar{\nu}_e H$ & $H \to \tau^+\tau^-$ \\
$e^+ e^- \to b\bar{b}H$ & $H \to \tau^+\tau^-$ \\
\hline
\end{tabular}
\end{minipage}%
\hfill
\begin{minipage}[t]{0.48\linewidth}
\centering
\textbf{$\sqrt{s}=365$~GeV}

\vspace{0.5em}

\begin{tabular}{ll}
\hline
Process & Decay \\
\hline
$e^+ e^- \to W^+ W^-$ & Inclusive $WW$ \\
$e^+ e^- \to ZZ$ & Inclusive $ZZ$ \\
$e^+ e^- \to t\bar{t}$ & Inclusive $t\bar{t}$ \\
$e^+ e^- \to \tau^+ \tau^-$ & Inclusive $\tau^+\tau^-$ \\
$e^+ e^- \to \nu_e \bar{\nu}_e H$ & $H \to \tau^+\tau^-$ \\
$e^+ e^- \to b\bar{b}H$ & $H \to \tau^+\tau^-$ \\
\hline
\end{tabular}
\end{minipage}

\end{table*}

Each background sample contains at least $10^5$ unscaled events. The targeted integrated luminosity for FCC-ee Higgs-strahlung run is 2.7~ab$^{-1}$ at $\sqrt{s} = 240$~GeV and 0.67 ab$^{-1}$ at $\sqrt{s} = 365$~GeV~\cite{FeasibilityReport}.  The detector response is modeled using the fast simulation software Delphes with the IDEA FCC-ee card~\cite{delphes}. A \texttt{k4simDelphes} wrapper is used to convert Delphes objects to EDM4HEP format~\cite{Madlener2021}~\cite{Gaede2026_EDM4hep}. The analysis is performed using the \texttt{FCCAnalyses} software framework, which is based on \texttt{RDataFrame}~\cite{FCCAnalyses}. Finally, the \texttt{Combine} tool \cite{combine} is used to compute expected exclusion limits and discovery reach for long-lived staus at FCC-ee.

Previous searches in the maximally mixed scenario exclude staus with masses of approximately $126\text{--}260~\mathrm{GeV}$ for a proper decay length of $c\tau_0=50~\mathrm{mm}$, while proper decay lengths of approximately $21\text{--}94~\mathrm{mm}$ are excluded for $m_{\tilde{\tau}}=200~\mathrm{GeV}$ \cite{mykyta}. To explore regions beyond these limits, we consider staus with proper decay lengths of $50~\mathrm{cm}$--$50~\mathrm{m}$, with the accessible mass range set by the center-of-mass energy: $100\text{--}119~\mathrm{GeV}$ at $\sqrt{s}=240~\mathrm{GeV}$ and $120\text{--}180~\mathrm{GeV}$ at $\sqrt{s}=365~\mathrm{GeV}$.

\section{Signature Reconstruction}

A charged LLP can give rise to different experimental signatures depending on its decay length and decay products. If the LLP decays too close to the interaction point for its daughter particles to be resolved from the primary vertex, the event is effectively prompt and falls outside the scope of this study. For sufficiently long decay lengths, the LLP can travel a measurable distance before decaying inside the tracking volume, allowing either the LLP trajectory, the charged daughter trajectory, or both to be reconstructed. While our use case are long-lived staus, the following applies to the generic decay
\begin{equation}
\tilde{X}^{\pm}_{\mathrm{LLP}} \rightarrow X^\pm Y^0,
\end{equation}
where $X^\pm$ is a charged particle and $Y^0$ is a neutral particle.

The possible reconstruction topologies are illustrated in Figure~\ref{fig:llp_generic_signatures}. If the LLP decays before leaving sufficient hits to reconstruct its trajectory, its charged daughter tracks can nevertheless be reconstructed and traced back to a secondary vertex, provided the decay produces more than one charged particle. This topology gives rise to a \textbf{displaced-vertex (DV) signature}.

For larger decay lengths~(around 1~m), the LLP can leave sufficient hits for its trajectory to be reconstructed before the decay. If the charged daughter is also reconstructed, the parent and daughter trajectories meet at the decay position with a change in direction, producing a visible \textbf{kinked-track signature}. The kink therefore provides a characteristic signature of a charged LLP decaying inside the tracking volume.

At even larger decay lengths~($>5$~m), an increasing fraction of LLPs traverse the tracking volume without decaying. These \textbf{detector-stable} LLPs can be identified through their charged-particle trajectory and, for sufficiently slow particles, their delayed time of flight. The measured momentum and time of flight can then be used to determine the LLP velocity and estimate its mass.

Thus, the three signatures considered in this analysis (displaced vertices, kinked tracks, and detector-stable charged particles) provide complementary sensitivity to charged LLPs over a broad range of decay lengths.

\input{escaping_clean}

\paragraph{Tracking and timing assumptions}

The analysis uses the charged-particle tracking provided by the FCC-ee detector
simulation in \textsc{Delphes}, with a dedicated modification to account for
metastable charged particles. In the standard tracking implementation, a
charged particle that decays sufficiently far from the interaction point may
not be retained as a reconstructed track. To account for this effect for the
long-lived stau benchmark, we implemented a metastable-particle tracking
procedure that allows staus to be reconstructed as tracks until their
decay point. The tracking performance is based on the assumed IDEA detector
geometry and parametrized tracking response, with the drift chamber providing
the main tracking volume.

Since \textsc{Delphes} provides track parameters but not individual hits, hits
are emulated for each reconstructed track. The trajectory is propagated through
the detector in a helix approximation, with the sensitive layers modeled as
cylinders in the barrel and disks in the endcaps, and each crossing of such a
surface is recorded as a hit at the crossing position. For the metastable staus,
the helix ends at the decay point, and the daughter track starts there.  From these hits, the total number of hits, the number of
drift-chamber hits, and the positions of the first and last hits are determined.
They are used to identify the characteristic hit patterns of kinked-track
topologies and to apply the corresponding hit-count and hit-veto requirements.


For the timing-based reconstruction, the LLP is assumed to originate from the
interaction point at the time of the collision. The time of flight is measured at
the first timing-layer surface reached by the particle, with a Gaussian timing
resolution of $30$~ps. The curvature of the trajectory in the magnetic field is
neglected in the path length, which is a good approximation for the
high-momentum staus considered here, since the resulting correction is
well below the timing resolution. The bunch spacing at both center-of-mass
energies is 25~ns, and is much larger than the arrival time of even the
slowest signal staus at the timing layer, so the measured time can be associated
unambiguously with the corresponding bunch crossing.



\paragraph{Reconstruction}

A \textbf{displaced vertex} topology is only relevant for LLP decays producing multiple charged particles. DV candidates are reconstructed by fitting at least two tracks that are displaced from the primary interaction point to a common vertex. In signal decays, the gravitino and neutrinos carry away a large fraction of the momentum, so the momentum of the visible tracks is not aligned with the displacement vector from the primary vertex to the reconstructed vertex. This is quantified by the pointing angle $\alpha_{\mathrm{DV}}$ between the vertex momentum and this displacement vector, and $\cos\alpha_{\mathrm{DV}}<0.95$ is required, which rejects vertices whose momentum and displacement are aligned, as expected for SM decays of long-lived hadrons such as $K_S$ and $\Lambda$.



A \textbf{kinked topology} is identified by pairing an incoming and an outgoing track with the same electric charge and requiring a minimum angular separation of $\Delta R > 0.05$ between them. The selected track pair is then fitted to a common vertex, which defines the kinked vertex (KV). The $\Delta R$ requirement is motivated by the detector geometry and ensures that the two tracks are sufficiently separated to be reconstructed as distinct tracks.

A hit-veto requirement based on the expected decay geometry is also imposed. Since the LLP does not produce hits beyond its decay position and the daughter particle does not produce hits before the decay, the hit pattern of the incoming and outgoing tracks is required to be consistent with this topology, thereby suppressing combinations of unrelated tracks. The two tracks are each required to have at least ten hits in total, including at least eight hits in the drift chamber, which improves track and vertex quality while reducing combinatorial backgrounds. Since in-flight decays and multiple interactions of charged particles with detector material are not explicitly modeled in Delphes and can produce kinks in the tracks, an additional requirement is imposed on the kink angle. The kink angle is defined as the angle between the incoming LLP trajectory and the outgoing daughter trajectory. Only kink vertices with a kink angle larger than~$20$~degrees are retained. This is expected to fully eliminate the non-simulated background.

For a sufficiently massive LLP, the particle is produced with a low velocity, particularly when its mass approaches the kinematic limit for production. If the particle is also sufficiently long-lived, it can traverse the tracking volume without decaying. Such a heavy, slowly moving LLP (with $\beta <1$, where $\beta=\frac{v}{c}$) is therefore expected to arrive at the timing layer later than relativistic SM particles. This delayed arrival can be exploited through a time-of-flight (TOF) measurement. Using the particle momentum reconstructed from the track and the known path length to the timing layer, its mass can be estimated as
\begin{equation}
    m_{\mathrm{esc}}
    =
    \frac{p_T}{\sin\theta}
    \sqrt{\frac{\mathrm{TOF}^2}{L^2}-1},
\end{equation}
where $p_T$ is the transverse momentum, $\theta$ is the polar angle, and $L$ is the path length from the interaction point to the timing layer.
\input{escaping}

\begin{figure*}[t]
    \centering
    \includegraphics[width=0.75\linewidth]{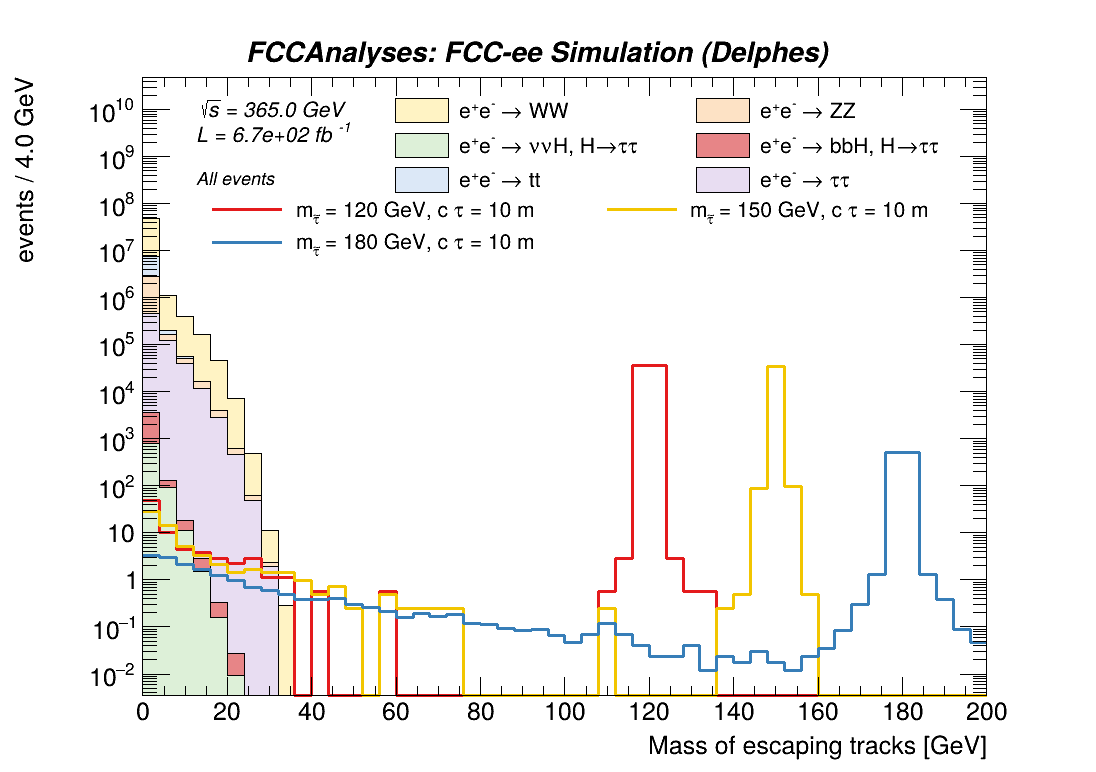}
    \caption{Reconstructed mass of escaping tracks obtained from the
    TOF measurement for signal benchmark points and for SM background
    processes. The signal peaks near the true stau mass, whereas the background populates the low-mass region.}
    \label{fig:Mass_of_escaping_track}
\end{figure*}

The path length is determined by the first timing-layer surface reached by the particle and is therefore given by
\begin{equation}
    L =
    \min\left(
        \frac{R}{|\sin\theta|},
        \frac{R'}{|\cos\theta|}
    \right),
\end{equation}
where $R$ is the radial distance to the transverse timing layer and $R'$ is the longitudinal distance to the endcap, as illustrated in Figure~\ref{fig:combined}. The corresponding detector dimensions are taken from the Delphes detector configuration. The reconstructed mass for several signal benchmark scenarios is shown in Figure~\ref{fig:Mass_of_escaping_track}. This signature is categorized as the \textbf{detector-stable LLP}~(DS) region.

\section{Stau Search}

Depending on the decay modes of the two \(\tau\) leptons in the benchmark scenario, the signal is divided into two categories for LLPs decaying inside the detector. Events are required to contain no more than one reconstructed electron or muon. In addition, the total number of reconstructed tracks is required to be fewer than nine, reflecting the low track multiplicity expected for the signal.

The detailed selections are as follows. Events are first classified according to their vertex topology:

\begin{itemize}
\item \textbf{Displaced-vertex topology:} Require no kinked vertices and
one or two displaced vertices satisfying the DV reconstruction requirements
described above. In the benchmark stau scenario, the gravitino is nearly massless and can carry away a large fraction of the available energy. Consequently, the invariant mass of the visible particles associated with the DV is expected to be relatively small, and an upper requirement of $m_{\mathrm{inv}}<40$~GeV is imposed.
\item \textbf{Kinked-vertex topology:} Require at least one kinked vertex satisfying the hit-pattern requirements described above.
\end{itemize}

To ensure that the kinked and displaced-vertex categories are mutually exclusive, DV candidates satisfying the hit-pattern requirements of the kinked topology are excluded.  The above topologies are then divided into two channels according to the lepton multiplicity of the final state:

\begin{itemize}
\item \textbf{Semi-leptonic selection:} Require exactly one reconstructed electron or muon with transverse momentum $p_T > 10~\mathrm{GeV}$.
\item \textbf{Hadronic selection:} Require zero reconstructed electrons or muons.
\end{itemize}

For the \textbf{detector-stable~(DS)} staus, we require no KVs or DVs but the mass of a primary track~$>35$~GeV.

These selections define five mutually exclusive signal regions, allowing them to be statistically combined. 
The leptonic decay of both \(\tau\) leptons is not explicitly targeted in the signal-region definition because of its relatively small branching fraction. Such events are nevertheless expected to produce two kinked vertices accompanied by two leptons. The combined set of kinked-vertex, displaced-vertex, and detector-stable selections provides sensitivity a large range of stau decay lengths. The same detector-stable-particle strategy can, in principle, be extended to larger lifetimes. 

\section{Exclusion Limits and Discovery Reach}

The \textsc{combine}~\cite{combine} tool is used to perform a statistical counting experiment combining the KV, DV and DS signal regions. Each simulated event is weighted according to
\begin{equation}
w = \frac{\sigma \mathcal{L}}{N_{\mathrm{MC}}},
\end{equation}
where \(\sigma\) is the production cross section, \(\mathcal{L}\) the integrated luminosity, and \(N_{\mathrm{MC}}\) the number of generated events.

No background MC events survive the KV and detector-stable selections. For the kinked-vertex selection, this is consistent with the strong suppression achieved by the applied requirements, in particular the kink-angle selection, which is expected to remove backgrounds from detector effects not modeled in \textsc{Delphes}. For the detector-stable selection, an independent full-detector-simulation study of long-lived staus at a 10~TeV muon collider likewise finds that no simulated background events survive its final selection~\cite{muoncolliderGMSBstau}. Since the signature relies on the same kind of observables, namely track quality, timing, and a reconstructed mass in a more challenging environment, we expect the \textsc{Delphes}-based MC to give a similar performance as the full simulation, which supports our zero-background result.

\begin{figure*}[t]
    \centering
     \includegraphics[width=0.75\linewidth]{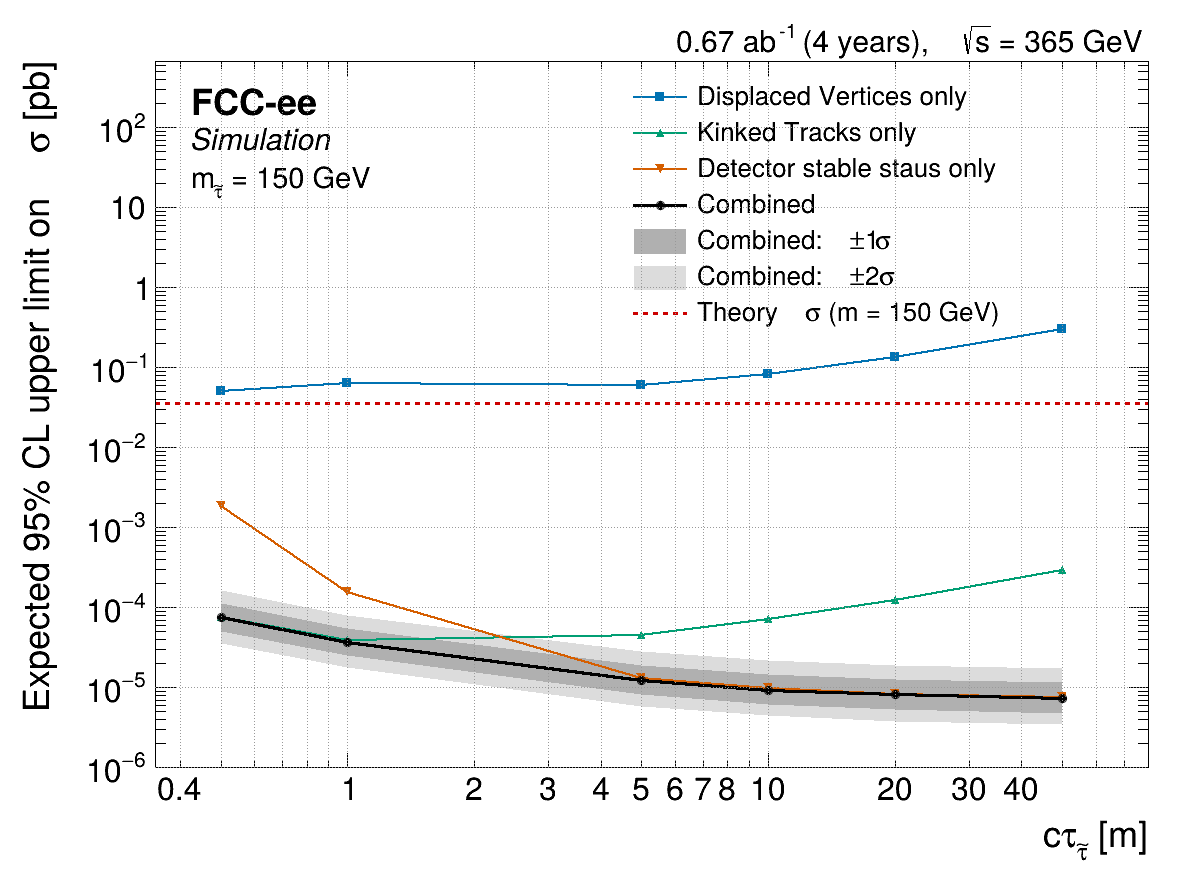}
    \caption{Expected 95\% CL upper limits on the stau pair-production cross section
    as a function of the stau proper decay length $c\tau$, for $m_{\tilde{\tau}} = 150$~GeV
    at $\sqrt{s} = 365$~GeV and an integrated luminosity of 0.67~ab$^{-1}$.
    The limits obtained using only the displaced-vertex (DV) regions, only the
    kinked-track (KV) regions, and only the escaping-stau regions are compared with
    the combined analysis which uses all three topologies. The gray bands show the
    $\pm1\sigma$ and $\pm2\sigma$ uncertainty ranges of the nominal expected limit.
    Only the luminosity uncertainty is included in the limit calculation.
    The dashed red line indicates the theoretical cross section.}
    \label{fig:limit_onemass}
\end{figure*}
Since a finite MC sample cannot establish a vanishing background rate, a conservative upper bound is assigned to each background sample with no surviving events. For zero surviving events, the one-sided 95\% CL upper limit on the Poisson mean is approximately three events, corresponding to
\begin{equation}
B < \frac{3}{N_{\mathrm{MC}}}\,\sigma\,\mathcal{L},
\end{equation}
with $\sigma$ and $N_{\mathrm{MC}}$ now referring to the respective background process. 

Figure~\ref{fig:limit_onemass} compares the expected 95\% CL cross-section limits obtained with each of the three reconstruction strategies and for their combination, for $m_{\tilde{\tau}} = 150$~GeV. The three strategies provide complementary sensitivity across the lifetime range. At short lifetimes, most staus decay close to the interaction point, leaving too few hits on the incoming track for a kink to be reconstructed. Their decay products can still be reconstructed as a displaced vertex, so the DV regions retain most of the signal. Because these regions are exposed to SM backgrounds, which are not suppressed as strongly as in the other regions, their sensitivity is nevertheless limited. At intermediate lifetimes, the KV regions become the most sensitive, as the staus traverse a substantial part of the tracking volume before decaying and leave enough hits for the kink to be reconstructed. For even larger decay lengths, most staus leave the tracker before decaying, and the escaping-stau regions provide the best sensitivity.

The combined analysis is at least as sensitive as the best single strategy at every lifetime and improves on it most strongly at the transitions between regimes. Comparing with the theoretical cross section, indicated by the dashed line, the combined limit lies below the prediction over the whole range of lifetimes.

\begin{figure*}
\centering
\includegraphics[width=0.85\linewidth]{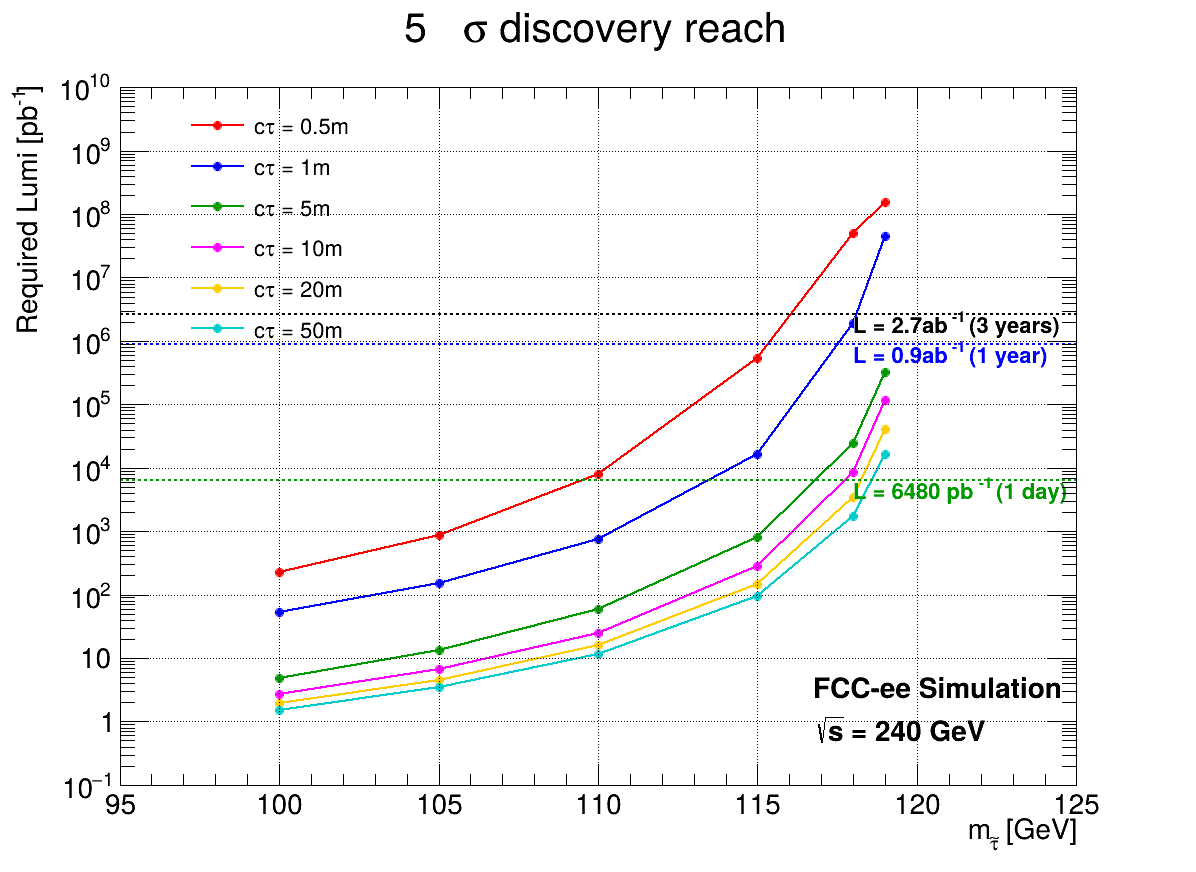}

```
\vspace{0.3cm}

\includegraphics[width=0.85\linewidth]{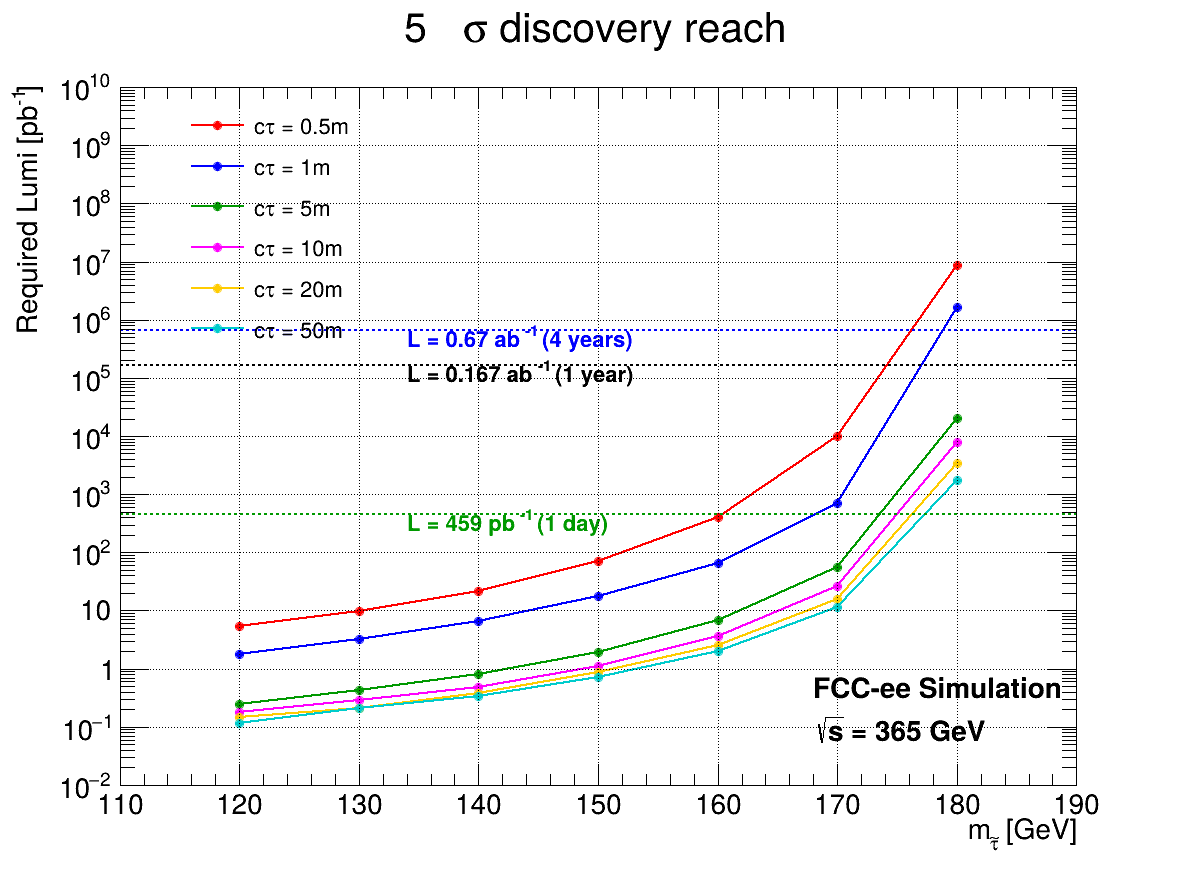}

\caption{Integrated luminosity required for a $5\sigma$ discovery as a function of stau mass
for different proper lifetime hypotheses at the FCC-ee, for \(\sqrt{s}=240\)~GeV and
\(\sqrt{s}=365~\mathrm{GeV}\). The horizontal lines indicate the integrated luminosity
delivered by the FCC-ee in one day, one year, and three to four years of data taking at
the respective center-of-mass energy. Most of the considered parameter space can be
probed with less than one year of data taking.}
\label{fig:discovery}

\end{figure*}

Figure~\ref{fig:discovery} shows the integrated luminosity required to achieve a 5~$\sigma $ discovery as a function of stau mass for the considered lifetime hypotheses. The required luminosity increases towards shorter lifetimes due to the reduced signal efficiency and towards larger masses due to the decreasing production cross section near the kinematic threshold. At $\sqrt{s} = 240$~GeV, stau masses below approximately 115~GeV can reach 5~$\sigma $ significance with relatively short data-taking periods over a broad range of lifetimes, while the near-threshold points at 118 and 119~GeV require substantially larger luminosities for less favorable lifetime hypotheses. At  $\sqrt{s} = 365$~GeV, a similar lifetime dependence is observed, including increased luminosity requirements for masses close to the kinematic threshold. At both center-of-mass energies, however, most of the considered parameter space can reach 5~$\sigma $ significance with less than one year of FCC-ee running.

 \section{Summary and Outlook}

We have studied the sensitivity of the IDEA detector at FCC-ee to charged long-lived particles (LLPs), using the long-lived stau in the Gauge Mediated Supersymmetry Breaking~(GMSB) model as a benchmark. The analysis combines kinked-track, displaced-vertex, and detector-stable-particle signatures for the first time, to probe a broad range of LLP lifetimes at FCC-ee. The results show that the clean environment and precise tracking and timing of FCC-ee make it well suited to discovering charged LLPs, with most of the considered parameter space reaching $5\sigma$ discovery significance in less than one year of data taking.

The three signatures are complementary across the lifetime range. Displaced-vertex reconstruction provides sensitivity at shorter lifetimes. As the lifetime increases, the particles traverse a larger fraction of the tracking volume before decaying, and kinked-track reconstruction becomes increasingly important; here the large tracking volume of IDEA provides substantial acceptance and holds great potential for the reconstruction of kinked tracks. For sufficiently long lifetimes, the LLPs escape the tracking volume before decaying and are instead identified through the detector-stable topology.

The discovery reach is evaluated at both $\sqrt{s}=240$~GeV and $\sqrt{s}=365$~GeV. At $240$~GeV, staus with masses below approximately $115$~GeV can reach $5\sigma$ significance with relatively short data-taking periods over a broad range of lifetimes, while near-threshold masses require larger integrated luminosities. A similar lifetime dependence is observed at $365$~GeV, with sensitivity extending to higher stau masses. At both center-of-mass energies, most of the considered parameter space can be discovered with less than one year of FCC-ee running. The sensitivity decreases near the kinematic limit because of the falling production cross section, and at short lifetimes because of the reduced efficiency for reconstructing displaced signatures. Future studies incorporating a full detector simulation is needed to confirm that backgrounds from detector effects not modeled in \textsc{Delphes} are negligible.

Although the benchmark considered in this study is the GMSB stau, the kinked-track and detector-stable topologies are more general and can be used to probe charged LLP decays of the form
\[
X^\pm_{\mathrm{LLP}} \rightarrow \ell^\pm + X^0,
\]
where $\ell$ denotes a charged particle and $X^0$ is a neutral particle. The same topologies can therefore arise in a range of models beyond GMSB. In particular, long-lived staus can also occur in supersymmetric scenarios such as the MSSM, as discussed in~\cite{japan_gmsb}. While no model-specific MSSM limits are derived here, the reconstruction strategy could be applied to such scenarios and to other charged LLP models. Future studies incorporating a full detector simulation and additional LLP production and decay modes can further establish the reach of FCC-ee for this broader class of signatures.

\paragraph{Implications for detector design}
Since the feasibility of these searches is driven by the reconstruction of slow and delayed tracks, our results are particularly relevant at the present detector design stage of FCC-ee. Slow, massive LLPs arrive late at the tracker and the timing layers, so the detector performance for such particles directly determines the reach. We therefore identify three requirements, which would benefit LLP searches at FCC-ee in general, not only the study presented here.

First, the tracking system must record and reconstruct late-arriving tracks. This requires sufficiently long readout windows and dedicated reconstruction for such tracks, for which standalone drift chamber tracking would be a significant asset.

Second, efficient displaced-vertex and kinked-track reconstruction requires tracking that does not assume tracks originate from the interaction point. The tracking must remain efficient for displaced tracks, tracks without silicon hits, and tracks with missing inner or outer hits. It must also provide good vertex resolution and efficiency for low-multiplicity vertices located far from the beam axis in the drift chamber.

Third, precise timing is essential for detecting and characterizing charged LLPs. The timing detector outside the tracker volume identifies slow and delayed particles and, through the measured momentum, allows the LLP mass to be reconstructed, which is key for characterizing detector-stable particles in the event of a discovery. The 30~ps resolution assumed in the Delphes simulation of IDEA works well for this analysis and serves as a benchmark to be maintained in the final detector. 

\bibliography{reference}
\end{document}

%% file: abstract.tex
Long-lived particles~(LLPs) have emerged as a compelling signature of physics beyond the standard model, offering unique discovery opportunities at current and future colliders. We present an analysis of charged LLPs leading to signatures with kinked tracks and displaced vertices. Using timing information the search is extended to detector-stable LLPs. We explore the discovery prospects of such particles at the Future Circular Collider~(FCC-ee), a proposed  $e^{+}e^{-}$ collider. The clean collision environment of FCC-ee and excellent tracking performance of the IDEA detector enhance the sensitivity to kinked-track signatures, which are considerably more challenging to reconstruct in the high-occupancy environment of hadron colliders. The excellent timing resolution of the timing layer enables us to infer the mass of the detector stable particles. As a benchmark the stau pair production in the gauge mediated supersymmetry breaking~(GMSB) within the minimal supersymmetric standard model scenario is considered. In this model the lightest supersymmetric particle is a gravitino and the next-to-lightest supersymmetric particle is the stau, the superpartner of the tau lepton. Discovery sensitivity up to the kinematic limit is obtained for the $ZH$ and $t\bar{t}$ runs of the planned FCC-ee operation. We demonstrate that a potential discovery would allow for a precise $\tilde\tau$ mass measurement.

%% file: escaping_clean.tex
\begin{figure*}[t]
\centering

\begin{minipage}[t]{0.30\linewidth}
\centering
\begin{tikzpicture}[scale=0.9]
  \colorlet{IP}{red}
  \colorlet{llpcol}{orange!95!black}
  \colorlet{pioncol}{red!85!black}
  \colorlet{invcol}{black!50}
  \colorlet{dvcol}{blue!70!black}

  \useasboundingbox (-0.3,-1.7) rectangle (3.9,2.6);

  \coordinate (IP) at (0,0);
  \coordinate (DV) at (1.8,0.9);

  \fill[IP] (IP) circle (2pt);
  \node[IP, anchor=north east, font=\scriptsize] at (-0.05,-0.05) {IP};

  \draw[llpcol, very thick, -{Latex[length=7pt, width=3.5pt]}]
    (IP) -- (DV);
  \node[llpcol, anchor=south, font=\small]
    at (0.8,0.55) {LLP};

  \fill[dvcol] (DV) circle (2.2pt);

  \draw[invcol, thick, dashed, -{Latex[length=10pt]}]
    (DV) -- ++(95:1.4) node[anchor=south, font=\small] {inv.};

  \draw[pioncol, thick, -{Latex[length=5pt]}]
    (DV) -- ++(35:1.7)  node[anchor=west, font=\small]  {$X^{\pm}$};
  \draw[pioncol, thick, -{Latex[length=5pt]}]
    (DV) -- ++(-15:1.6) node[anchor=north west, font=\small] {$X^{\mp}$};
  \draw[pioncol, thick, -{Latex[length=5pt]}]
    (DV) -- ++(-60:1.5) node[anchor=north, font=\small] {$X^{\pm}$};

\end{tikzpicture}
\\[2pt]
{\small\bfseries Displaced vertex}
\end{minipage}
\hfill
\begin{minipage}[t]{0.30\linewidth}
\centering
\begin{tikzpicture}[scale=0.9]
  \colorlet{IP}{red}
  \colorlet{llpcol}{orange!95!black}
  \colorlet{pioncol}{red!85!black}
  \colorlet{invcol}{black!50}
  \colorlet{SVcol}{blue}
  \colorlet{anglecol}{blue!60!black}

  \useasboundingbox (-0.3,-1.7) rectangle (4.2,2.6);

  \coordinate (IP) at (0,0);
  \coordinate (KV) at (1.8,1.04);  

  \fill[IP] (IP) circle (2pt);
  \node[IP, anchor=north east, font=\scriptsize] at (-0.05,-0.05) {IP};

  \draw[llpcol, very thick, -{Latex[length=7pt, width=3.5pt]}]
    (IP) -- (KV);
  \node[llpcol, anchor=south, font=\small]
    at (0.8,0.55) {LLP};

  \fill[SVcol] (KV) circle (2.2pt);

  \draw[llpcol, thick, dash pattern=on 20pt off 3pt, opacity=0.6]
    (KV) -- ++(30:2.2);

  \draw[pioncol, thick, -{Latex[length=5pt]}]
    (KV) -- ++(0:1.9) node[anchor=west, font=\small] {$X^{\pm}$};

  \draw[invcol, thick, dashed, -{Latex[length=10pt]}]
    (KV) -- ++(95:1.4) node[anchor=south, font=\small] {inv.};

  \draw[anglecol, thick]
    (KV) ++(30:1.0) arc[start angle=30, end angle=0, radius=1.0];

  \node[anglecol, font=\scriptsize, anchor=south west, inner sep=1pt]
    at ([shift={(1.2,0.25)}]KV) {$\alpha_{\text{kink}}$};

\end{tikzpicture}
\\[2pt]
{\small\bfseries Kinked track}
\end{minipage}
\hfill
\begin{minipage}[t]{0.30\linewidth}
\centering
\begin{tikzpicture}[scale=0.9]
  \colorlet{IP}{red}
  \colorlet{llpcol}{orange!95!black}
  \colorlet{pioncol}{red!85!black}
  \colorlet{invcol}{black!50}
  \colorlet{detcol}{black!55}
  \colorlet{dvcol}{blue!70!black}
  \colorlet{anglecol}{blue!60!black}

  \useasboundingbox (-1.6,-1.7) rectangle (4.3,2.6);

  \coordinate (IP) at (0,0);

  \draw[detcol, thick] (-1.5,-1.3) rectangle (1.5,1.3);
  \node[detcol, font=\scriptsize\bfseries, anchor=north]
    at (0,-1.35) {Detector};

  \fill[IP] (IP) circle (2pt);
  \node[IP, anchor=north east, font=\scriptsize] at (-0.05,-0.05) {IP};

  \coordinate (EXIT) at (1.5,0.65);
  \coordinate (DV) at (2.3,1.0);

  \draw[llpcol, very thick, -{Latex[length=7pt, width=3.5pt]}]
    (IP) -- (DV);
  \node[llpcol, anchor=south, font=\small] at (0.75,0.30) {LLP};

  \fill[detcol] (EXIT) circle (1.8pt);

  \fill[dvcol] (DV) circle (2.2pt);

  \draw[pioncol, thick, -{Latex[length=5pt]}]
    (DV) -- ++(-1.5:1.5) node[anchor=west, font=\small] {$X^{\pm}$};

  \draw[invcol, thick, dashed, -{Latex[length=10pt]}]
    (DV) -- ++(80:1.2) node[anchor=south, font=\small] {inv.};

\end{tikzpicture}
\\[2pt]
{\small\bfseries Detector-stable LLP}
\end{minipage}

\caption{Generic signatures of a long-lived charged particle (LLP) that we aim to exploit: a displaced vertex with multiple charged particles, a kinked track from a decay with an invisible daughter, and an LLP that escapes the detector volume before decaying.}
\label{fig:llp_generic_signatures}
\end{figure*}

%% file: escaping.tex
\begin{figure*}[t]
\centering
\begin{minipage}{0.85\linewidth}
\centering
\begin{tikzpicture}[scale=1.0]
  \colorlet{beamcol}{black}
  \colorlet{SVcol}{blue}
  \colorlet{Bccol}{blue}
  \colorlet{IP}{red}
  \colorlet{pioncol}{red!90!black}
  \colorlet{pioncol1}{red!85!black}
  \colorlet{staucol1}{orange!95!black}
  \colorlet{staucol2}{orange!95!black}
  \colorlet{invcol}{black!50}
  \colorlet{invcol1}{black!50}
  \colorlet{dvcol}{blue!70!black}
  \colorlet{timingcol}{green!55!black}
  \colorlet{geomcol}{blue!60!black}

  \coordinate (IP) at (0,0);

  \fill[timingcol!12] (-3.0,-2.4) rectangle (3.0, 2.4);
  \draw[timingcol, line width=1.8pt] (-3.0,-2.4) rectangle (3.0, 2.4);
  \node[timingcol, font=\small\bfseries, anchor=south west]
    at (3.05, 2.4) {Timing Layer};

  \draw[geomcol, thin, {Latex[length=5pt]}-{Latex[length=5pt]}]
    (0.0, 0.05) -- (0.0, 2.35);
  \node[geomcol, anchor=west, font=\small] at (-0.12, 1.2) {$R$};

  \draw[geomcol, thin, {Latex[length=5pt]}-{Latex[length=5pt]}]
    (0.05, -0.32) -- (2.95, -0.32);
  \node[geomcol, anchor=north, font=\small] at (1.5, -0.36) {$R'$};

  \draw[beamcol, very thick, dashed, -{Latex[length=8pt]}] (-6.2,0) -- (0,0);
  \draw[beamcol, very thick, dashed, -{Latex[length=8pt]}] (+6.2,0) -- (0,0);
  \node[beamcol, anchor=north] at (-5.2,-0.18) {$e^+$};
  \node[beamcol, anchor=north] at ( 5.2,-0.18) {$e^-$};

  \fill[IP, draw=IP, thick] (IP) circle (2.3pt);
  \node[IP, anchor=north east, font=\small] at (-0.12,-0.12) {IP};

  \coordinate (TLexit) at (1.62, 2.4);
  \coordinate (KV) at (2.05, 3.05);

  \draw[staucol2, very thick, -{Latex[length=8pt, width=4pt]}]
    (IP) -- (KV);

  \draw[black, thin] (0.7, 0) arc[start angle=0, end angle=56, radius=0.7];
  \node[black, font=\small] at (0.90, 0.40) {$\theta$};

  \draw[geomcol, very thick, -{Latex[length=7pt, width=4pt]}]
    (TLexit) -- ++(90:1.0);
  \node[geomcol, anchor=east, font=\small] at (1.55, 3.05) {$p_T$};

  \node[staucol2, anchor=west, font=\small] at (1.50, 2.10) {$\tilde{\tau}^{-}$};
  \node[staucol2, anchor=west, font=\small] at (1.30, 1.65) {$(\beta{<}1)$};

  \fill[timingcol!80!black] (TLexit) circle (3.5pt);
  \node[timingcol!70!black, font=\scriptsize, anchor=west, align=left]
    at (0.2, 3.15) {TOF hit\\$t_{\tilde{\tau}^{-}}$};

  \fill[SVcol, draw=Bccol, thick] (KV) circle (2.5pt);

  \draw[staucol2, thin, dashed, opacity=0.35]
    (KV) -- ++(56:0.7);

  \draw[pioncol, line width=1.3pt, -{Latex[length=6pt]}]
    (KV) -- ++(15:2.8)
    node[anchor=west] {$\pi^{-}$};

  \draw[invcol, line width=1pt, dashed, -{Latex[length=6pt]}]
    (KV) -- ++(100:2.0)
    node[anchor=south] {$\tilde{G}$};

  \coordinate (TLexit2) at (-1.62, -2.4);
  \coordinate (DV) at (-2.05, -3.05);

  \draw[staucol1, very thick, -{Latex[length=8pt, width=4pt]}]
    (IP) -- (DV);
  \node[staucol1, anchor=east, font=\small]
    at (-0.85, -1.1) {$\tilde{\tau}^{+}$};
  \node[staucol1, anchor=east, font=\small]
    at (-0.55, -0.55) {$(\beta < 1)$};

  \fill[timingcol!80!black] (TLexit2) circle (3.5pt);
  \node[timingcol!70!black, font=\scriptsize, anchor=east, align=right]
    at (0.5, -3.15) {TOF hit\\$t_{\tilde{\tau}^{+}}$};

  \fill[dvcol] (DV) circle (2.5pt);

  \draw[pioncol1, line width=1.2pt, -{Latex[length=6pt]}]
    (DV) -- ++(-80:2.2)  node[anchor=north]      {$\pi^\pm$};
  \draw[pioncol1, line width=1.2pt, -{Latex[length=6pt]}]
    (DV) -- ++(-130:2.0) node[anchor=north east] {$\pi^\mp$};
  \draw[pioncol1, line width=1.2pt, -{Latex[length=6pt]}]
    (DV) -- ++(-165:1.9) node[anchor=east]       {$\pi^\pm$};

  \draw[invcol1, line width=1pt, dashed, -{Latex[length=6pt]}]
    (DV) -- ++(-20:2.8)
    node[anchor=north west] {$\tilde{G}$};

\end{tikzpicture}
\end{minipage}
\caption{Signal topology and mass-reconstruction geometry for the detector-stable signature. Both $\tilde{\tau}$ tracks ($\beta < 1$) cross the timing layer and register a delayed TOF hit ($t_{\tilde{\tau}^{-}}$ and $t_{\tilde{\tau}^{+}}$), from which the mass is reconstructed. The decays that occur beyond the timing layer are not used. For the $\tilde{\tau}^{\pm}$, the track length $L$ to the timing layer is set by whichever wall is reached first: the transverse wall at distance $R$ or the endcap at distance $R'$ along the beam axis $z$, with $\theta$ measured from the beam line and $p_T$ the transverse momentum at the hit. The decay modes are drawn to show how the same benchmark populates the other two signatures when the lifetime is short enough for the decay to occur inside the tracking volume: $\tilde{\tau}^{-} \to \pi^{-} + \tilde{G}$ would lead to a kinked track, and $\tilde{\tau}^{+} \to 3\pi + \tilde{G}$ would produce a displaced vertex. }
\label{fig:combined}
\end{figure*}

%% file: reference.bib
@article{GMSB_basic,
  author  = {Giudice, G. F. and Rattazzi, R.},
  title   = {Theories with {G}auge-{M}ediated {S}upersymmetry {B}reaking},
  journal = {Physics Reports},
  year    = {1998},
  volume  = {322},
  pages   = {419--499},
  doi     = {10.1016/S0370-1573(99)00042-3},
  eprint  = {hep-ph/9801271},
  archivePrefix = {arXiv},
  primaryClass  = {hep-ph}
}

@article{GMSB_ee,
  author  = {Ambrosanio, Sandro and Blair, Grahame A.},
  title   = {{M}easuring {G}auge-{M}ediated {S}upersymmetry {B}reaking {P}arameters at a 500 {GeV} {$e^+e^-$} {L}inear {C}ollider},
  journal = {The European Physical Journal C},
  year    = {1999},
  eprint  = {hep-ph/9905403},
  archivePrefix = {arXiv},
  primaryClass  = {hep-ph}
}

@article{japan_gmsb,
  author  = {Jittoh, Toshifumi and Sato, Joe and Shimomura, Takashi and Yamanaka, Masato},
  title   = {{L}ong {L}ife {S}tau in the {M}inimal {S}upersymmetric {S}tandard {M}odel},
  journal = {Physical Review D},
  volume  = {73},
  number  = {5},
  year    = {2006},
  doi     = {10.1103/PhysRevD.73.055009}
}

@article{madgraph,
  author  = {Alwall, J. and Frederix, R. and Frixione, S. and Hirschi, V. and Maltoni, F. and Mattelaer, O. and Shao, H.-S. and Stelzer, T. and Torrielli, P. and Zaro, M.},
  title   = {The {A}utomated {C}omputation of {T}ree-{L}evel and {N}ext-to-{L}eading {O}rder {D}ifferential {C}ross {S}ections, and their {M}atching to {P}arton {S}hower {S}imulations},
  journal = {Journal of High Energy Physics},
  year    = {2014},
  volume  = {2014},
  number  = {7},
  pages   = {079},
  doi     = {10.1007/JHEP07(2014)079}
}

@article{combine,
  author  = {Hayrapetyan, A. and Tumasyan, A. and Adam, W. and others},
  title   = {The {CMS} {S}tatistical {A}nalysis and {C}ombination {T}ool: {C}ombine},
  journal = {Computing and Software for Big Science},
  volume  = {8},
  number  = {1},
  year    = {2024},
  doi     = {10.1007/s41781-024-00121-4}
}

@article{pythia8,
  author  = {Sjöstrand, Torbjörn and Ask, Stefan and Christiansen, Jesper R. and Corke, Richard and Desai, Nishita and Ilten, Philip and Mrenna, Stephen and Prestel, Stefan and Rasmussen, Christine O. and Skands, Peter Z.},
  title   = {An {I}ntroduction to {PYTHIA} 8.2},
  journal = {Computer Physics Communications},
  volume  = {191},
  pages   = {159--177},
  year    = {2015},
  doi     = {10.1016/j.cpc.2015.01.024}
}

@misc{mykyta,
  author        = {{CMS Collaboration}},
  title         = {{S}earch for the {P}air {P}roduction of {L}ong-{L}ived {S}upersymmetric {P}artners of the {T}au {L}epton in {P}roton-{P}roton {C}ollisions at {$\sqrt{s}=13$} {TeV}},
  year          = {2026},
  eprint        = {2601.17576},
  archivePrefix = {arXiv},
  primaryClass  = {hep-ex},
  url           = {https://arxiv.org/abs/2601.17576}
}

@misc{FeasibilityReport,
  author    = {Benedikt, Michael and Bartmann, Wolfgang and Burnet, Jean-Paul and Carli, Christian and Chance, Antoine and Craievich, Paolo and Giovannozzi, Massimo and Grojean, Christophe and Gutleber, Johannes and Hanke, Klaus and Henriques, Andre and Janot, Patrick and Lourenco, Carlos and Mangano, Michelangelo and Otto, Thomas and Poole, John Howard and Rajagopalan, Srini and Raubenheimer, Tor and Todesco, Ezio and Ulrici, Luisa and Watson, Timothy Paul and Wilkinson, Guy and Zimmermann, Frank},
  title     = {{F}uture {C}ircular {C}ollider {F}easibility {S}tudy {R}eport {V}olume 2: {A}ccelerators, {T}echnical {I}nfrastructure and {S}afety},
  year      = {2025},
  publisher = {CERN Document Server},
  doi       = {10.17181/CERN.EBAY.7W4X},
  url       = {https://cds.cern.ch/record/2928793}
}

@article{LLP_alimena,
  author  = {Alimena, Juliette and Beacham, James and Borsato, Martino and Cottin, Giovanna and Curtin, David and Knapen, Simon and Kraml, Sabine and Liu, Zhen and Ramsey-Musolf, Michael J. and Shelton, Jessie and Shuve, Brian and Verducci, Monica and Zurita, Jose and Adams, Todd and Alpigiani, Cristiano and Apresyan, Artur and Buchmueller, Oliver and Buschmann, Malte and Campanelli, Mario and Coccaro, Andrea and Conte, Eric and Craig, Nathaniel and Dall'Occo, Elena and De Simone, Andrea and Deppisch, Frank F. and Dienes, Keith R. and Dildick, Sven and Drewes, Marco and Galon, Iftah and Gershtein, Yuri and Giammanco, Andrea and Golling, Tobias and Haas, Andrew and Helo, J. C. and Hesketh, Gavin and Hirsch, Martin and Ilten, Philip and Knapen, Simon and Kraml, Sabine and Landsberg, Greg and Lubatti, Henry and Mariotti, Alberto and Moortgat, Filip and Nachman, Benjamin and Pascoli, Silvia and Pinfold, James L. and Schuster, Philip and Schwaller, Pedro and Stolarski, Daniel and Trocino, Daniele and Trovato, Marco and Tsai, Yuhsin and Walker, Devin G. E. and Young, Charles and Yu, Jiang-Hao and Zalewski, Piotr and others},
  title   = {{S}earching for {L}ong-{L}ived {P}articles beyond the {S}tandard {M}odel at the {L}arge {H}adron {C}ollider},
  journal = {Journal of Physics G: Nuclear and Particle Physics},
  volume  = {47},
  number  = {9},
  pages   = {090501},
  year    = {2020},
  doi     = {10.1088/1361-6471/ab4574}
}

@misc{delphes,
  author        = {Ovyn, S. and Rouby, X. and Lemaitre, V.},
  title         = {{DELPHES}, a {F}ramework for {F}ast {S}imulation of a {G}eneric {C}ollider {E}xperiment},
  year          = {2010},
  eprint        = {0903.2225},
  archivePrefix = {arXiv},
  primaryClass  = {hep-ph},
  url           = {https://arxiv.org/abs/0903.2225}
}

@article{GMSBllp,
  author  = {Dimopoulos, Savas and Dine, Michael and Raby, Stuart and Thomas, Scott},
  title   = {{E}xperimental {S}ignatures of {L}ow {E}nergy {G}auge-{M}ediated {S}upersymmetry {B}reaking},
  journal = {Physical Review Letters},
  volume  = {76},
  number  = {19},
  pages   = {3494--3497},
  year    = {1996},
  doi     = {10.1103/PhysRevLett.76.3494}
}

@misc{theideastudygroup2025ideadetectorconceptfccee,
      title={The IDEA detector concept for FCC-ee}, 
      author={The IDEA Study Group},
      year={2025},
      eprint={2502.21223},
      archivePrefix={arXiv},
      primaryClass={physics.ins-det},
      url={https://arxiv.org/abs/2502.21223}, 
}

@misc{Madlener2021,
  author    = {Madlener, Thomas and Völkl, Valentin and Helsens, Clément and Gaede, Frank and Chrząszcz, Marcin},
  title     = {{key4hep/k4SimDelphes}: {Z}enodo {R}elease},
  year      = {2021},
  publisher = {Zenodo},
  doi       = {10.5281/zenodo.4564683}
}

@misc{FCCAnalyses,
  author       = {{HEP-FCC Collaboration}},
  title        = {{FCCAnalyses}},
  year         = {2024},
  howpublished = {GitHub repository},
  url          = {https://github.com/HEP-FCC/FCCAnalyses}
}

@software{Gaede2026_EDM4hep,
  author    = {Gaede, Frank and Madlener, Thomas and Sailer, André Philippe and Völkl, Valentin and Carceller, Juan Miguel and Fila, Mateusz Jakub},
  title     = {{EDM4hep}},
  year      = {2026},
  publisher = {Zenodo},
  doi       = {10.5281/zenodo.18310626}
}

@article{DM_evidence_1,
  author  = {Clowe, Douglas and Bradač, Marusa and Gonzalez, Anthony H. and Markevitch, Maxim and Randall, Scott W. and Jones, Christine and Zaritsky, Dennis},
  title   = {A {D}irect {E}mpirical {P}roof of the {E}xistence of {D}ark {M}atter},
  journal = {The Astrophysical Journal},
  volume  = {648},
  number  = {2},
  pages   = {L109--L113},
  year    = {2006},
  doi     = {10.1086/508162}
}

@incollection{MARTIN_1998,
       author = {{Martin}, Stephen P.},
        title = "{A Supersymmetry Primer}",
    booktitle = {Perspectives on Supersymmetry. Edited by KANE GORDON L. Published by World Scientific Publishing Co. Pte. Ltd},
         year = 1998,
       editor = {{Kane}, Gordon L.},
        pages = {1-98},
          doi = {10.1142/9789812839657_0001},
       adsurl = {https://ui.adsabs.harvard.edu/abs/1998pesu.book....1M}
}

@article{Particle_physics_review,
  author        = {Navas, S. and Amsler, C. and Gutsche, T. and Hanhart, C. and Hernandez-Rey, J. J. and Lourenco, C. and Masoni, A. and Mikhasenko, M. and Mitchell, R. E. and Patrignani, C. and Schwanda, C. and Spanier, S. and Venanzoni, G. and Yuan, C. Z. and others},
  collaboration = {Particle Data Group},
  title         = {{R}eview of {P}article {P}hysics},
  journal       = {Physical Review D},
  volume        = {110},
  number        = {3},
  pages         = {030001},
  year          = {2024},
  doi           = {10.1103/PhysRevD.110.030001}
}

@article{Asai2011,
  author        = {Asai, Shoji and Azuma, Yuya and Endo, Motoi and Hamaguchi, Koichi and Iwamoto, Sho},
  title         = {{S}tau {K}inks at the {LHC}},
  journal       = {Journal of High Energy Physics},
  volume        = {2011},
  number        = {12},
  pages         = {041},
  year          = {2011},
  doi           = {10.1007/JHEP12(2011)041},
  eprint        = {1103.1881},
  archivePrefix = {arXiv},
  primaryClass  = {hep-ph}
}

@article{Klamka2026,
  author        = {Klamka, Jan and {\.Z}arnecki, Aleksander Filip},
  title         = {{S}earching for {L}ong-{L}ived {P}articles with the {ILD} {E}xperiment},
  journal       = {European Physical Journal Plus},
  volume        = {141},
  number        = {5},
  pages         = {481},
  year          = {2026},
  doi           = {10.1140/epjp/s13360-026-07701-w},
  eprint        = {2605.03908},
  archivePrefix = {arXiv},
  primaryClass  = {hep-ex}
}

@article{Benjamin_Fuks,
   title={{S}imulating spin-{$\frac{3}{2}$} {P}articles at {C}olliders},
   volume={73},
   ISSN={1434-6052},
   url={http://dx.doi.org/10.1140/epjc/s10052-013-2580-x},
   DOI={10.1140/epjc/s10052-013-2580-x},
   number={10},
   journal={The European Physical Journal C},
   publisher={Springer Science and Business Media LLC},
   author={Christensen, N. D. and de Aquino, P. and Deutschmann, N. and Duhr, C. and Fuks, B. and Garcia-Cely, C. and Mattelaer, O. and Mawatari, K. and Oexl, B. and Takaesu, Y.},
   year={2013},
   month=Oct }

@misc{muoncolliderGMSBstau,
      title={{E}nabling {S}earches for {L}ong-{L}ived {P}articles at a {F}uture 10 {TeV} {M}uon {C}ollider}, 
      author={Mira Littmann and Mark Larson and Kate Wandrisco and Benjamin Rosser and Tate Flicker and Kane Huang and Leo Rozanov and Karri Folan Di Petrillo},
      year={2026},
      eprint={2512.10097},
      archivePrefix={arXiv},
      primaryClass={hep-ex},
      url={https://arxiv.org/abs/2512.10097}, 
}
